\documentclass[preprint,preprintnumbers, prd, floatfix, superscriptaddress,nofootinbib] {revtex4-1}
\usepackage{epsfig}
\usepackage{subfigure}
\usepackage{dcolumn}
\usepackage{bm}
\usepackage[usenames ,dvipsnames]{xcolor}
\usepackage{slashed}
\usepackage{graphicx,color}
\begin{document}
\title{Testing Non-Standard Neutrinos in Purely Leptonic Lepton Decays }

\author{Han Zhang}
\affiliation{School of Physics, Zhengzhou University, Zhengzhou 450001, People's Republic of China}
\author{Bai-Cian Ke}
\email{Corresponding author: baiciank@ihep.ac.cn}
\affiliation{School of Physics, Zhengzhou University, Zhengzhou 450001, People's Republic of China}
\author{Yao Yu}
\email{Corresponding author: yuyao@cqupt.edu.cn}
\affiliation{Chongqing University of Posts \& Telecommunications, Chongqing, 400065, China}
 \affiliation{Department of Physics and Chongqing Key Laboratory for Strongly Coupled Physics, Chongqing University, Chongqing 401331, People's Republic of China}

\date{\today}

\begin{abstract}
  We propose a method to probe sterile neutrinos using polarization observables
  in the purely leptonic decays
  $\ell^{\prime-} \to \ell^{-} \bar{\nu}_{\ell} \nu_{\ell^{\prime}}$. By
  analyzing angular distributions and asymmetries derived from polarized decay
  rates, we identify distinctive signatures of sterile neutrino mixing. In
  particular, we demonstrate that sterile neutrinos can induce singularities
  in certain asymmetry parameters as functions of the invariant mass squared of
  the neutrino pair. These singularities occur for sterile neutrino masses
  $m_{4\nu}$ satisfying $m_{4\nu}^2 < m_{\ell^{\prime}}^2 / 2$, providing a
  clear target for experimental investigation. Our results motivate the
  incorporation of polarized beam sources at future colliders to enhance
  sensitivity to sterile neutrinos and other new physics.
\end{abstract}

\maketitle
\section{introduction}
The observation of neutrino
oscillations~\cite{Davis:1968cp,Super-Kamiokande:1998kpq,SNO:2002tuh,KamLAND:2002uet,K2K:2002icj},
which demonstrate that neutrinos have mass and undergo lepton flavor mixing,
stands as one of the most significant discoveries of physics beyond the
original Standard Model~(SM). Decades of neutrino oscillation experiments have
established a consistent framework of neutrino masses and mixing that
successfully explains observations from solar, atmospheric, reactor, and
accelerator experiments. The neutrino oscillation anomalies observed in these
measurements further suggest the possible existence of additional neutrino
states beyond the three well-known neutrino mass eigenstates. Such new states
would need to be predominantly electroweak singlets, making them "sterile"
neutrinos~\cite{Marquez:2022bpg,Beuthe:2001rc,Dasgupta:2021ies} that would mix
with the active neutrinos while interacting only very weakly through SM forces.
In the presence of sterile neutrinos, the mixing between flavor eigenstates
$\nu_\ell$~($\ell = e,\, \mu,\, \tau$) and the mass eigenstates $\nu_i$ is
described by a generalized lepton mixing matrix $U_{\ell i}$:
\begin{equation}
\nu_\ell = \sum_{i=1}^{3+n_s} U_{\ell i} \nu_i,
\end{equation}
where $n_s$ denotes the number of sterile
neutrinos~\cite{Bryman:2019ssi,Bryman:2019bjg}. The existence of sterile
neutrinos constitutes one of the most pressing unresolved questions in modern
particle physics. This fundamental mystery drives significant experimental and
theoretical efforts to precisely constrain their possible mixing
parameters~(or coupling) $|U_{\ell i}|^2$.

Current studies about sterile neutrinos primarily target neutrino oscillation,
neutrinoless double beta decay, branching ratios of leptonic decays. In this
work, we investigate, as a complementary signature, angular distribution of
polarization observables in purely leptonic decays of leptons, induced by the
mixing of a massive sterile neutrino with the active neutrino sector. We
analyze the decays
$\ell^{\prime-}\to \ell^-\bar{\nu}_\ell\nu_{\ell^{\prime}}$, where
$(\ell^{\prime}=\tau\,,\ell=e,\mu)$ and $(\ell=\mu\,,\ell=e)$,
and compute the corrections to SM predictions arising from sterile neutrinos,
with a particular focus on their impact on angular distribution of
polarization observables. The sterile neutrino mass is considered across over
a range from $\sim \! \mathrm{MeV}$ to $\mathrm{GeV}$, with the assumption of
$n_s = 1$ sterile state, denoted $\nu_4$; generalization to $n_s \geq 2$ is
straightforward. A heavy $\nu_4$ in this mass range is unstable and therefore
avoids the stringent cosmological constraint on the sum of stable neutrino
masses, $\sum_i m_{\nu_i} \lesssim 0.12~\mathrm{eV}$\cite{Planck:2018vyg}. We
restrict our attention to treat $\nu_4$ as a Dirac fermion, which automatically
satisfy current constraints from neutrinoless double beta
decays~\cite{CUORE:2017tlq,GERDA:2018pmc}.

\section{Formalism}
The decay rate of $\ell^{\prime-}\to \ell^-\bar{\nu}_\ell\nu_{\ell^{\prime}}$
in the rest frame of $\ell^{\prime}$ can be written in terms of an invariant
amplitude ${\cal M}$
\begin{eqnarray}\label{eqz1}
  \Gamma&=&\frac{(2\pi)^4}{2m_{\ell^{\prime}}}\int\frac{d^{3}p_{\ell}d^{3}p_{i}d^{3}p_{j}}{(16\pi^{3})^3E_{\ell}E_{i}E_{j}}|{\cal M}(\ell^{\prime-}\to \ell^-\bar{\nu}_i\nu_{j})|^2\delta^{4}(p_{\ell}+p_{i}+p_{j}-p_{\ell^{\prime}})\equiv\frac{1}{2^{9}\pi^{5}m_{\ell^{\prime}}}{\cal I}\,.
  \label{eq:Gamma}
\end{eqnarray}
Here, $m_{\ell^{\prime}}$ is the rest mass of $\ell^{\prime}$, and
$p_{\ell^{\prime}, \ell, i, j}$ ($E_{\ell^{\prime}, \ell, i, j}$) are the
four-momenta (energy) of $\ell^{\prime}$, $\ell$, $\nu_{i}$, and $\nu_{j}$ in
the rest frame of $\ell^{\prime}$, respectively. The subscript $i, j$ denotes
the mass eigenstates of neutrinos. The $\delta$ function
$\delta^{4}(p_{\ell}+p_{i}+p_{j}-p_{\ell^{\prime}})$ enforces the conservation
of four-momentum between the parent and the final-state particles.

\begin{figure}[t!]
  \centering
  \includegraphics[width=4.0in]{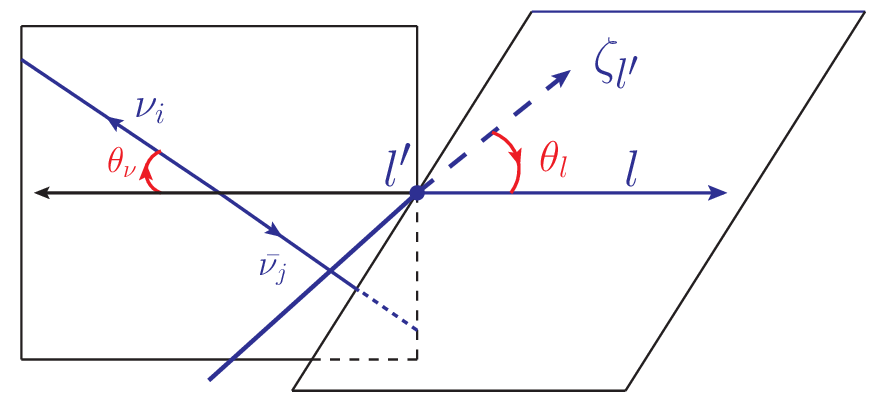}
  \caption{
    Kinematics of the decay
    $\ell^{\prime-}\to \ell^-\bar{\nu}_\ell\nu_{\ell^{\prime}}$. The initial
    charged lepton $\ell^{\prime}$ has polarization $\zeta_{\ell^{\prime}}$.
    The angle $\theta_\ell$ is defined in the $\ell$ rest frame as the angle
    between the $\ell$ three-momentum and $\zeta_{\ell^{\prime}}$. The angle
    $\theta_\nu$ is defined in the $\bar{\nu}_\ell\nu_{\ell^{\prime}}$
    center-of-mass frame as the angle between the three-momenta of
    $\bar{\nu}_\ell$ and $\nu_{\ell^{\prime}}$.
  }\label{tu1}
\end{figure}

Neutrinos are undetectable in conventional detectors. However, the total
four-momentum of the $\nu_{i}\nu_{j}$ system can be inferred through
four-momentum conservation. Experimentally, it can be reconstructed by
subtracting the measured four-momentum of the final-state charged lepton from
the known initial four-momentum, such as $\mu$ and $\tau$. This
``missing four-momentum'' technique has been widely employed in leptonic
decay analyses, for example, in measurements of
$D\to \ell\nu_\ell$~\cite{Ke:2023qzc}. Accordingly, we introduce a $Y$-frame
representing the center-of-mass frame of the $\nu_{i}\nu_{j}$ system and
${\cal I}$ can be rewritten by recursively factorizing the $\delta$ function
as~\cite{Cabibbo:1965zzb}

\begin{eqnarray}\label{eqz2}
  {\cal I} &=& \int\frac{d^{3}p_{\ell}d^{3}p_{i}d^{3}p_{j}}{E_{\ell}E_{i}E_{j}}|{\cal M}(\ell^{\prime-}\to \ell^-\bar{\nu}_i\nu_{j})|^2\delta^{4}(p_{\ell}+p_{i}+p_{j}-p_{\ell^{\prime}})\nonumber\\
  &=&\int \frac{d^{3}p_{i}}{E_{i}}\frac{d^{3}p_{j}}{E_{j}}\delta^{4}(p_i+p_j-Y)
  \cdot\int\frac{d^{3}p_{\ell}}{E_{\ell}}d^{4}Y\delta^{4}(p_\ell+Y-p_{\ell^{\prime}})|{\cal M}(\ell^{\prime-}\to \ell^-\bar{\nu}_i\nu_{j})|^2\,.
\end{eqnarray}
The integrations over the $\delta$ functions can be carried out independently:
\begin{eqnarray}\label{eqt2}
  \int \frac{d^{3}p_{i}}{E_{i}}\frac{d^{3}p_{j}}{E_{j}}\delta^{4}(p_i+p_j-Y) &=& \frac{|\vec{p}_{i}|}{\sqrt{Y^2}}d\Omega_{\nu}\,,
\end{eqnarray}
where $\Omega_\nu$  and $\vec{p}_{i}$ represent the solid angle and the
three-momentum of $\nu_i$ in the $Y$-frame, respectively;
\begin{eqnarray}\label{eqt3}
\int\frac{d^{3}p_{\ell}}{E_{\ell}}d^{4}Y\delta^{4}(p_\ell+Y-p_{\ell^{\prime}}) &=&  dY^2\frac{|\vec{Y}|}{2m_{\ell^{\prime}}}d\Omega_{\ell}\,,
\end{eqnarray}
where the solid angle $\Omega_{\ell}$ and three-momentum
$\vec{p}_\ell=-\vec{Y}$ are defined in the $\ell^{\prime}$ rest frame, and the
polarization of $\ell^{\prime}$~($\zeta_{\ell^{\prime}}$) is chosen as the
$z$-axis, as shown in Fig.~\ref{tu1}.

By substituting Eqs.~(\ref{eqz2})-(\ref{eqt3}) into Eq.~(\ref{eqz1}), we obtain
the expression for the differential decay width with respect to polarization:
\begin{eqnarray}
  \frac{d\Gamma_{\lambda\lambda^\prime}}{dY^2d\cos\theta_\ell}&=&\frac{1}{2^{10}\pi^{5}m^2_{\ell^{\prime}}}\int|{\cal M_{\lambda\lambda^\prime}}(\ell^{\prime-}\to \ell^-\bar{\nu}_i\nu_{j})|^2\frac{|\vec{p}_{i}|}{\sqrt{Y^2}}|\vec{Y}|d\Omega_{\nu}d\phi_{\ell}\,,
  \label{eq:Gamma2}
\end{eqnarray}
where $\lambda^{(\prime)}$ represent the polarizations of $\ell^{(\prime)}$.
Since the polarization and flavor of neutrinos are unmeasurable in detectors,
while the polarization of other related charged leptons can, in principle, be
accessed experimentally, we will sum over the neutrino polarizations and
integrate out the associated angular variables in the calculation of the decay
amplitude. The leptonic weak decay amplitude for
$\ell^{\prime-} \to \ell^- \bar{\nu}_j \nu_i$ can be rewritten using a Fierz
rearrangement~\cite{Shrock:1980ct,Shrock:1981wq,Shrock:1981cq}, which
separates measurable from unmeasurable parameters and given by:
\begin{eqnarray}\label{amp1a}
{\cal M_{\lambda\lambda^\prime}}(\ell^{\prime-}\to \ell^-\bar{\nu}_j\nu_{i})&=&\frac{G_F}{\sqrt 2}U_{\ell j}U^*_{\ell^{\prime}i}
\bar{u}_\ell(\lambda)\gamma^\mu(1-\gamma^5)u_{\ell^\prime}(\lambda^\prime)\bar{u_i}\gamma_\mu(1-\gamma^5)\nu_j\,.
\end{eqnarray}
The lepton mixing matrix $U_{\ell j}$ is a $3 \times 3$ unitary matrix within
the SM, while it is a $4 \times 4$ unitary matrix taking into account the
sterile neutrino. Note that neutrinos are assumed to be of the Dirac type. Then
the squared magnitude is given by
\begin{eqnarray}\label{eqt4}
  |{\cal M}(\ell^{\prime-}\to \ell^-\bar{\nu}_j\nu_{i})|^2 &=&\sum_{i,j} \frac{G_F^2}{2}|U_{\ell j}|^2|U^*_{\ell^{\prime}i}|^2T(\lambda,\lambda^\prime,m_i,m_j)\,.
\end{eqnarray}
Here, $T(\lambda,\lambda^\prime,m_i,m_j)$ encodes the dependence on the charged
lepton polarizations $\lambda^{(\prime)}$ and the neutrino masses $m_{i, j}$.
\begin{eqnarray}\label{ew1}
  T(\lambda,\lambda^\prime, m_i,m_j)&=&F(\lambda,\lambda^\prime,m)F^*(\lambda,\lambda^\prime,n)g_{mm'}g_{nn'}\epsilon^\dagger_{\mu}(m')\epsilon_{\nu}(n')L^{\mu\nu}
\end{eqnarray}
with
\begin{eqnarray}\label{L1}
  L^{\mu\nu}&=&\sum_{\lambda_{\nu}}\bar{u_i}\gamma^\mu(1-\gamma^5)\nu_j\bar{\nu_j}\gamma^\nu(1-\gamma^5)u_i\nonumber\\
  &=&-8g^{\mu\nu}p_i\cdot p_j+8(p_i^\mu p_j^\nu+p_i^\nu p_j^\mu)-8i\epsilon^{\mu\nu \alpha\beta }(p_i)_{\alpha} (p_j)_{\beta}
\end{eqnarray}
and
\begin{eqnarray}\label{eqb1}
  F(\lambda,\lambda^\prime,m)&=&\bar{u}_\ell(\lambda)\gamma_\mu(1-\gamma^5)u_{\ell^\prime}(\lambda^\prime)\epsilon^\mu(m)\nonumber\\
  &=&\zeta_\ell^\dag(\lambda)\frac{p_\ell\cdot\sigma+m_\ell}{\sqrt{(p_\ell^0+m_\ell)}}
  \bar{\sigma}_{\mu}\frac{p_{\ell^\prime}\cdot\sigma+m_{\ell^\prime}}{\sqrt{(p_{\ell^\prime}^0+m_{\ell^\prime})}}\zeta_{\ell^\prime}(\lambda^\prime)\epsilon^\mu(m)\,,
\end{eqnarray}
where $\zeta(\lambda)$ is a two-component spinor. In the above, sets of
polarization vectors $\epsilon^\mu(m)$ are introduced to relate different rest
frames. These polarization vectors satisfy the orthonormality and completeness
properties given by
\begin{eqnarray}\label{eqc1}
  &&\epsilon^\dagger_{\mu}(m)\epsilon^{\mu}(n)=g_{mn}\,\,\,\,(m,n=t,\pm,0)\,,\nonumber\\
  &&\epsilon^\mu(m)\epsilon^{\dagger\nu}(n)g_{m n}=g^{\mu\nu}\,,
  \label{eq:orthonormality}
\end{eqnarray}
where
$g_{m n}={\rm diag}(+, -, -, -)={\rm diag}(g_{tt}, g_{++}, g_{--}, g_{00})$.
In the $\ell^{\prime}$ rest frame, one has
\begin{eqnarray}\label{eqa1}
  p^{\mu}_{\ell^{\prime}} &=& (m_{\ell^{\prime}},0,0,0)\,, \nonumber\\
  p^{\mu}_{\ell} &=&(p_\ell^0,|\vec{Y}|\sin\theta_\ell\cos\phi_\ell,|\vec{Y}|\sin\theta_\ell\sin\phi_\ell,|\vec{Y}|\cos\theta_\ell)\,, \nonumber\\
  Y^\mu &=&(p_Y^0,|\vec{Y}|\sin(\pi-\theta_\ell)\cos(\pi+\phi_\ell),|\vec{Y}|\sin(\pi-\theta_\ell)\sin(\pi+\phi_\ell),|\vec{Y}|\cos(\pi+\phi_\ell))\,.
\end{eqnarray}
The polarization vectors for the $Y$ system in the $\ell^{\prime}$ rest frame
can be written as
\begin{eqnarray}\label{eqa2}
  \epsilon^\mu(t) &=& \frac{1}{\sqrt{Y^2}}(p_Y^0,-|\vec{Y}|\sin\theta_\ell\cos\phi_\ell,-|\vec{Y}|\sin\theta_\ell\sin\phi_\ell,-|\vec{Y}|\cos\theta_\ell)\nonumber\\
  \epsilon^\mu(0) &=& \frac{1}{\sqrt{Y^2}}(|\vec{Y}|,-p_Y^0\sin\theta_\ell\cos\phi_\ell,-p_Y^0\sin\theta_\ell\sin\phi_\ell,-p_Y^0\cos\theta_\ell)\nonumber\\
  \epsilon^\mu(+) &=&-\frac{1}{\sqrt{2}} (0,\cos\theta_\ell\cos\phi_\ell-i\sin\phi_\ell,
  \cos\theta_\ell\sin\phi_\ell+i\cos\phi_\ell,-\sin\theta_\ell)\nonumber\\
  \epsilon^\mu(-) &=&\frac{1}{\sqrt{2}} (0,\cos\theta_\ell\cos\phi_\ell+i\sin\phi_\ell,
  \cos\theta_\ell\sin\phi_\ell-i\cos\phi_\ell,-\sin\theta_\ell)\,,
  \label{eq:K2Q}
\end{eqnarray}
and the two-component spinors are given by
\begin{eqnarray}\label{eqa3}
  \zeta_\ell^\dag(+) &=&\left(
  \begin{array}{cc}
    \cos(\frac{\theta_\ell}{2}),\, & \sin(\frac{\theta_\ell}{2})e^{-i\phi_\ell} \\
  \end{array}
  \right)\,,
  \zeta_\ell^\dag(-) =\left(
  \begin{array}{cc}
    -\sin(\frac{\theta_\ell}{2})e^{i\phi_\ell} ,\, & \cos(\frac{\theta_\ell}{2}) \\
  \end{array}
  \right)\,,\nonumber\\
  \zeta_{\ell^\prime}(+) &=& \left(
  \begin{array}{c}
    1 \\
    0 \\
  \end{array}
  \right)\,,
  \zeta_{\ell^\prime}(-) = \left(
  \begin{array}{c}
    0 \\
    1 \\
  \end{array}
  \right)\,.
\end{eqnarray}
When $\ell^\prime$ is fully polarized, we take
$\zeta_{\ell^\prime}(\lambda^\prime)=\zeta_{\ell^\prime}(+)$ for convenience
and without loss of generality.

Next, one can obtain the expression of $F(\lambda,\lambda^\prime,m)$ for
different polarizations by substituting Eqs.~(\ref{eqa1})-(\ref{eqa3}) into
Eq.~(\ref{eqb1}). The corresponding explicit forms will be given later in
Eq.~(\ref{eq:F_polarization}), where only the terms relevant to the decay
width are retained. Employing the relations
\begin{eqnarray}
  p_i\cdot\epsilon^\dagger(t) &=& p_i^0 ,\,\,p_i\cdot\epsilon^\dagger(0)=-|\vec{p_i}|\cos\theta_i,\,\,p_i\cdot\epsilon^\dagger(\pm)=\pm\frac{1}{\sqrt{2}}|\vec{p_i}|\sin\theta_ie^{\mp i\phi_i}\,,\nonumber\\
  p_j\cdot\epsilon^\dagger(t) &=& p_j^0 ,\,\,p_j\cdot\epsilon^\dagger(0)=|\vec{p_i}|\cos\theta_i,\,\,p_j\cdot\epsilon^\dagger(\pm)=\mp\frac{1}{\sqrt{2}}|\vec{p_i}|\sin\theta_i e^{\mp i\phi_i}\,.
\end{eqnarray}
along with Eq.~(\ref{eqc1}), these results are inserted into Eq.~(\ref{ew1})
to derive the expression of $T(\lambda,\lambda^\prime, m_i,m_j)$. Subsequently,
with $T(\lambda,\lambda^\prime, m_i,m_j)$ and Eq.~(\ref{eqt4}), we obtain the
explicit formula for the differential decay width from Eq.~(\ref{eq:Gamma2}),
given by
\begin{eqnarray}
  \frac{d\Gamma_{\lambda\lambda^\prime}}{dY^2d\cos\theta_\ell}&=&\sum_{i,j}|U_{\ell j}|^2|U^*_{\ell^{\prime}i}|^2\frac{G^2_F}{2^{10}\pi^{4}m^2_{\ell^{\prime}}}\frac{|\vec{p}_{i}|}{\sqrt{Y^2}}|\vec{Y}| w_{\lambda\lambda^\prime} \,,
  \label{eq:Gamma3}
\end{eqnarray}
\begin{eqnarray}
  w_{\lambda\lambda^\prime}&=&32\pi |F(\lambda,\lambda^\prime,t)|^2(2p_i^0p_j^0-p_i\cdot p_j)
 +32\pi |F(\lambda,\lambda^\prime,0)|^2(p_i\cdot p_j-\frac{2}{3}|\vec{p}_i|^2)\nonumber\\
 &&+32\pi |F(\lambda,\lambda^\prime,+)|^2(p_i\cdot p_j-\frac{2}{3}|\vec{p}_i|^2)
 +32\pi |F(\lambda,\lambda^\prime,-)|^2(p_i\cdot p_j-\frac{2}{3}|\vec{p}_i|^2)\,.
\end{eqnarray}
With the helping of
\begin{eqnarray}\label{eq:F_polarization}
  |F(+,+,t)|^2 &=&  |F(+,+,0)|^2=\frac{m_{\ell^{\prime}}}{m_{\ell}+p^0_\ell} \frac{(Y^0-|\vec{p}_\ell|)^2}{Y^2}(m_{\ell}-|\vec{p}_\ell|+p^0_\ell)^2(1+\cos\theta_\ell)\nonumber\\
  |F(+,+,+)|^2&=&\frac{2m_{\ell^{\prime}}}{m_{\ell}+p^0_\ell} (m_{\ell}-|\vec{p}_\ell|+p^0_\ell)^2(1-\cos\theta_\ell),\,\,\, |F(+,+,-)|^2=0\nonumber\\
  |F(-,+,t)|^2 &=&  |F(-,+,0)|^2=\frac{m_{\ell^{\prime}}}{m_{\ell}+p^0_\ell} \frac{(Y^0+|\vec{p}_\ell|)^2}{Y^2}(m_{\ell}+|\vec{p}_\ell|+p^0_\ell)^2(1-\cos\theta_\ell)\nonumber\\
  |F(-,+,+)|^2&=&0,\,\,|F(-,+,-)|^2=\frac{2m_{\ell^{\prime}}}{m_{\ell}+p^0_\ell} (m_{\ell}+|\vec{p}_\ell|+p^0_\ell)^2(1+\cos\theta_\ell)\,,
\end{eqnarray}
$w_{\lambda\lambda^\prime}$ are expressed as
\begin{eqnarray}
  w_{++}&=&64\pi\frac{m_{\ell^{\prime}}}{Y^2}[(m^2_{\ell^{\prime}}+m^2_{\ell})p^0_\ell-2m_{\ell^{\prime}}m^2_{\ell}-(m^2_{\ell^{\prime}}-m^2_{\ell})|\vec{Y}|]  (2p_i^0p_j^0-\frac{2}{3}|\vec{p}_i|^2)(1+\cos\theta_\ell)\nonumber\\
 &&+128\pi m_{\ell^{\prime}}(p^0_\ell-|\vec{Y}|)(p_i\cdot p_j-\frac{2}{3}|\vec{p}_i|^2)(1-\cos\theta_\ell)\,,\nonumber\\
  w_{-+}&=&64\pi\frac{m_{\ell^{\prime}}}{Y^2}[(m^2_{\ell^{\prime}}+m^2_{\ell})p^0_\ell-2m_{\ell^{\prime}}m^2_{\ell}+(m^2_{\ell^{\prime}}-m^2_{\ell})|\vec{Y}|]  (2p_i^0p_j^0-\frac{2}{3}|\vec{p}_i|^2)(1-\cos\theta_\ell)\nonumber\\
 &&+128\pi m_{\ell^{\prime}}(p^0_\ell+|\vec{Y}|)(p_i\cdot p_j-\frac{2}{3}|\vec{p}_i|^2)(1+\cos\theta_\ell)\,,
\end{eqnarray}
with
\begin{eqnarray}\label{eq:p_i}
  |\vec{p}_i| &=&\frac{\sqrt{[Y^2-(m_i+m_j)^2][Y^2-(m_i-m_j)^2]}}{2\sqrt{Y^2}},\,\,p_i^0=\sqrt{|\vec{p}_i|^2+m_i^2}\,,\nonumber\\
  |\vec{Y}| &=&\frac{\sqrt{[m^2_{\ell^{\prime}}-(\sqrt{Y^2}+m^2_{\ell})^2][m^2_{\ell^{\prime}}-(\sqrt{Y^2}-m^2_{\ell})^2]}}{2m_{\ell^{\prime}}},\,\,p_\ell^0=\sqrt{|\vec{Y}|^2+m_\ell^2}\,.
\end{eqnarray}

Having obtained the expressions for the angular distributions corresponding to
different polarizations, we introduce a set of integrated asymmetries to
isolate and quantify potential new physics contributions. These asymmetries are
defined over specific angular domains as follows:
\begin{eqnarray}
  \int_{D_{1}}d\cos\theta_\ell \frac{d\Gamma_{+,+}}{dY^2d\cos\theta_\ell}&\equiv& A_1 ,\,\, \int_{D_{2}}d\cos\theta_\ell \frac{d\Gamma_{+,+}}{dY^2d\cos\theta_\ell}\equiv B_1\,,\nonumber\\
  \int_{D_{1}}d\cos\theta_\ell \frac{d\Gamma_{-,+}}{dY^2d\cos\theta_\ell}&\equiv& A_2 ,\,\, \int_{D_{2}}d\cos\theta_\ell \frac{d\Gamma_{-,+}}{dY^2d\cos\theta_\ell}\equiv B_2\,,
\end{eqnarray}
where $\int_{D_{1(2)}}\equiv\int_{-1}^0\pm\int_0^1$. From these integrated
quantities, several asymmetry parameters are constructed:
\begin{eqnarray}
  \Upsilon_1 &\equiv& \frac{A_2-A_1}{A_1+A_2},\,\,\Upsilon_2 \equiv\frac{B_1+B_2}{A_1+A_2},\,\,\Upsilon_3 \equiv\frac{B_1-B_2}{A_1+A_2}\,.
\end{eqnarray}
In the SM~($i,j=1,2,3$), where neutrino masses can be neglected, it follows
from Eq.~(\ref{eq:p_i}) that $|p_i|=p_i^0=|p_j|=p_j^0=\sqrt{Y^2}/2$ and
$p_i\cdot p_j = 2|p_i||p_j|= Y^2$. Consequently, $w_{\lambda\lambda^\prime}$ in
Eq.~(\ref{eq:Gamma3}) becomes independent of $i$ and $j$. Using the unitarity
of the mixing matrix, $\sum_j |U_{lj}|^2 = \sum_i|U^*_{l^\prime i}|^2 = 1$, the
polarized differential decay rates are given by
\begin{eqnarray}\label{eq36}
  \frac{d\Gamma^S_{+,+}}{dY^2d\cos\theta_\ell}&=&\frac{G^2_F}{2^{5}\pi^{4}m_{\ell^{\prime}}}|\vec{Y}|\left\{[(m^2_{\ell^{\prime}}+m^2_{\ell})p^0_\ell-2m_{\ell^{\prime}}m^2_{\ell}-(m^2_{\ell^{\prime}}-m^2_{\ell})|\vec{Y}|]  \frac{1+\cos\theta_\ell}{3}\right.\nonumber\\
 &&+\left.\frac{2}{3}Y^2(p^0_\ell-|\vec{Y}|)(1-\cos\theta_\ell)\right\}\,,\nonumber\\
 \frac{d\Gamma^S_{-,+}}{dY^2d\cos\theta_\ell}&=&\frac{G^2_F}{2^{5}\pi^{4}m_{\ell^{\prime}}}|\vec{Y}|\left\{[(m^2_{\ell^{\prime}}+m^2_{\ell})p^0_\ell-2m_{\ell^{\prime}}m^2_{\ell}+(m^2_{\ell^{\prime}}-m^2_{\ell})|\vec{Y}|]  \frac{1-\cos\theta_\ell}{3}\right.\nonumber\\
 &&+\left.\frac{2}{3}Y^2(p^0_\ell+|\vec{Y}|)(1+\cos\theta_\ell)\right\}\,.
\end{eqnarray}
The corresponding SM asymmetry parameters are then
\begin{eqnarray}\label{eq37}
  \Upsilon^S_1 &=& \frac{(m^2_{\ell^{\prime}}-m^2_{\ell})|\vec{Y}|+2Y^2|\vec{Y}|}{(m^2_{\ell^{\prime}}+m^2_{\ell})p^0_\ell-2m_{\ell^{\prime}}m^2_{\ell}+2Y^2p^0_\ell},\,\,\nonumber\\
  \Upsilon^S_2 &=&\frac{1}{2}\times \frac{2Y^2|\vec{Y}|-(m^2_{\ell^{\prime}}-m^2_{\ell})|\vec{Y}|}{(m^2_{\ell^{\prime}}+m^2_{\ell})p^0_\ell-2m_{\ell^{\prime}}m^2_{\ell}+2Y^2p^0_\ell},\,\,\nonumber\\
  \Upsilon^S_3 &=&\frac{1}{2}\times \frac{(m^2_{\ell^{\prime}}+m^2_{\ell})p^0_\ell-2m_{\ell^{\prime}}m^2_{\ell}-2Y^2p^0_\ell}{(m^2_{\ell^{\prime}}+m^2_{\ell})p^0_\ell-2m_{\ell^{\prime}}m^2_{\ell}+2Y^2p^0_\ell}\,.
\end{eqnarray}

In the presence of a sterile neutrino, one can derive asymmetry parameters
$\Upsilon^B_{1,2,3}$ having both the SM and new physics contributions from the
general decay rate expression in Eq.~(\ref{eq:Gamma3}) with additional mixing
matrix elements $i,j=4$ involved. The amplitude corresponding to the case where
both neutrinos are sterile mass eigenstates~($i=j=4$) involves the product
$U_{\ell 4}U^*_{\ell^{\prime}4}$, which is doubly suppressed by the small
active-sterile mixing and can therefore be safely neglected. In the following
discussion, we focus on the new physics contribution in which one of the
neutrinos is sterile while the other is a SM neutrino. As a result, the net
effect of sterile neutrinos reduces to the expression below:
\begin{eqnarray}
 \Delta\Upsilon_{1,2,3}&\equiv&\Upsilon^B_{1,2,3}-\Upsilon^S_{1,2,3}\,,
\end{eqnarray}
\begin{eqnarray}
 \Delta\Upsilon_1&\simeq&(U^2_{\ell^{\prime}4}+U^2_{\ell4})\left\{\Theta(Y^2-m^2_{4\nu})\right.\nonumber\\
 &&\times\left.[\frac{(1-\frac{m^4_{4\nu}}{Y^4})(m^2_{\ell^{\prime}}-m^2_{\ell})|\vec{Y}|+2(Y^2-m^2_{4\nu}\frac{2Y^2-m^2_{4\nu}}{Y^2})|\vec{Y}|}{(m^2_{\ell^{\prime}}+m^2_{\ell})p^0_\ell-2m_{\ell^{\prime}}m^2_{\ell}+2Y^2p^0_\ell}]\frac{2|\vec{p}_i|}{\sqrt{Y^2}}-\Upsilon^S_1\}\right.\,,\nonumber\\
 \Delta\Upsilon_2&\simeq&(U^2_{\ell^{\prime}4}+U^2_{\ell4})\left\{\Theta(Y^2-m^2_{4\nu})\right.\nonumber\\
 &&\times\left.[\frac{2(Y^2-m^2_{4\nu}\frac{2Y^2-m^2_{4\nu}}{Y^2})|\vec{Y}|-(1-\frac{m^4_{4\nu}}{Y^4})(m^2_{\ell^{\prime}}-m^2_{\ell})|\vec{Y}|}{(m^2_{\ell^{\prime}}+m^2_{\ell})p^0_\ell-2m_{\ell^{\prime}}m^2_{\ell}+2Y^2p^0_\ell}]\frac{|\vec{p}_i|}{\sqrt{Y^2}}-\Upsilon^S_2\}\right.\,,\nonumber\\
 \Delta\Upsilon_3&\simeq&(U^2_{\ell^{\prime}4}+U^2_{\ell4})\left\{\Theta(Y^2-m^2_{4\nu})\right.\nonumber\\
 &&\times\left.[\frac{(1-\frac{m^4_{4\nu}}{Y^4})[(m^2_{\ell^{\prime}}+m^2_{\ell})p^0_\ell-2m_{\ell^{\prime}}m^2_{\ell}]-2(Y^2-m^2_{4\nu}\frac{2Y^2-m^2_{4\nu}}{Y^2})p^0_\ell}{(m^2_{\ell^{\prime}}+m^2_{\ell})p^0_\ell-2m_{\ell^{\prime}}m^2_{\ell}+2Y^2p^0_\ell}]\frac{|\vec{p}_i|}{\sqrt{Y^2}}-\Upsilon^S_3\}\right.\,,
\end{eqnarray}
where $m_{4\nu}$ is the sterile neutrino mass and $\Theta(Y^2-m^2_{4\nu})$ is
Heaviside step function. For clarity in the subsequent discussion, we define
three normalized parameters that quantify the relative new physics corrections
to the Standard Model predictions. Each parameter is scaled by a factor of
$1/(U^2_{\ell^{\prime}4} + U^2_{\ell 4})$ to isolate the effect of the sterile
neutrino mixing.
\begin{eqnarray}
\delta\Upsilon_{\ell1}&\equiv& \frac{\Delta\Upsilon_1}{\Upsilon^S_1}\frac{1}{U^2_{\ell^{\prime}4}+U^2_{\ell4}}\simeq\Theta(Y^2-m^2_{4\nu})\nonumber\\
 &&\times[\frac{(1-\frac{m^4_{4\nu}}{Y^4})(m^2_{\ell^{\prime}}-m^2_{\ell})+2(Y^2-m^2_{4\nu}\frac{2Y^2-m^2_{4\nu}}{Y^2})}{(m^2_{\ell^{\prime}}-m^2_{\ell})+2Y^2}]\frac{2|\vec{p}_i|}{\sqrt{Y^2}}-1\,,\nonumber\\
\delta\Upsilon_{\ell2}&\equiv& \frac{\Delta\Upsilon_2}{\Upsilon^S_2}\frac{1}{U^2_{\ell^{\prime}4}+U^2_{\ell4}}\simeq\Theta(Y^2-m^2_{4\nu})\nonumber\\
 &&\times[\frac{2(Y^2-m^2_{4\nu}\frac{2Y^2-m^2_{4\nu}}{Y^2})-(1-\frac{m^4_{4\nu}}{Y^4})(m^2_{\ell^{\prime}}-m^2_{\ell})}{2Y^2-(m^2_{\ell^{\prime}}-m^2_{\ell})}]\frac{2|\vec{p}_i|}{\sqrt{Y^2}}-1\,,\nonumber\\
 \delta\Upsilon_{\ell3}&\equiv& \frac{\Delta\Upsilon_3}{\Upsilon^S_3}\frac{1}{U^2_{\ell^{\prime}4}+U^2_{\ell4}}\simeq\Theta(Y^2-m^2_{4\nu})\nonumber\\
 &&\times[\frac{(1-\frac{m^4_{4\nu}}{Y^4})[(m^2_{\ell^{\prime}}+m^2_{\ell})p^0_\ell-2m_{\ell^{\prime}}m^2_{\ell}]-2(Y^2-m^2_{4\nu}\frac{2Y^2-m^2_{4\nu}}{Y^2})p^0_\ell}{[(m^2_{\ell^{\prime}}+m^2_{\ell})p^0_\ell-2m_{\ell^{\prime}}m^2_{\ell}]-2Y^2p^0_\ell}]\frac{2|\vec{p}_i|}{\sqrt{Y^2}}-1\,.\nonumber\\
\end{eqnarray}
The SM provides definite predictions for the asymmetry
parameters~(Eq.~\ref{eq37}), which are derived from polarized decay rates.
Within the SM framework, the difference between these predictions and
experimental measurements is expected to be consistent with zero. Any
statistically significant deviation from zero at any values of $Y^2$ would
therefore suggest new physics beyond the SM, such as the existence of sterile
neutrinos. Note that $Y^2$ could be experimentally accessed through the
four-momentum different of measured final state lepton momentum and the known
initial four-momentum. All other neutrino-related degrees of freedom that are
not experimentally accessible have been fully integrated out at the theoretical
level. As a result, the predicted polarization-dependent observables are
expressed entirely in terms of experimentally measurable kinematics.

\section{Numerical Results and Discussions}
\begin{figure}[htbp]
\includegraphics[width=2.8in]{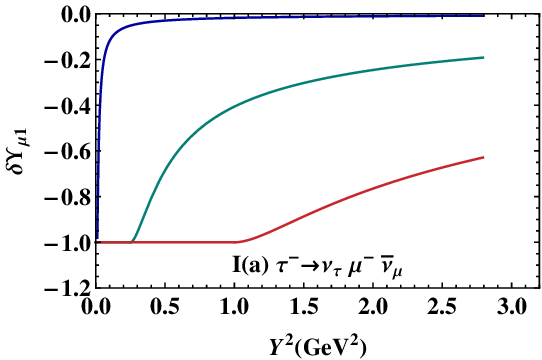}
\includegraphics[width=2.8in]{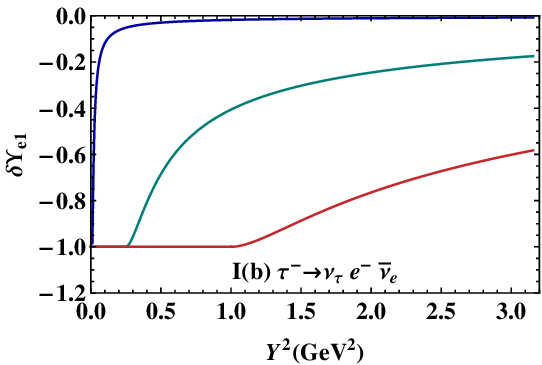}
\includegraphics[width=2.8in]{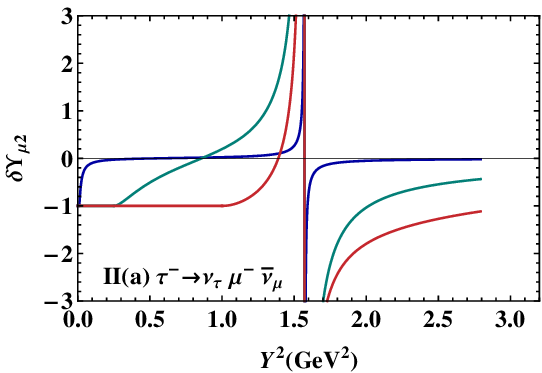}
\includegraphics[width=2.8in]{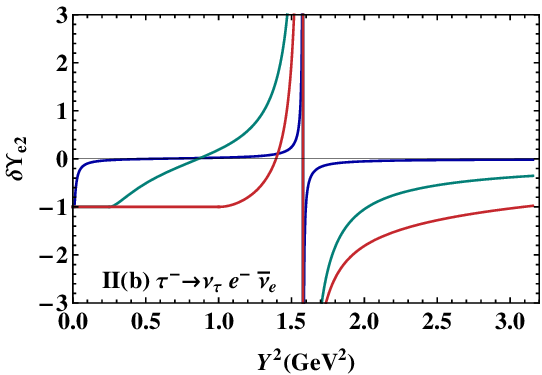}
\includegraphics[width=2.8in]{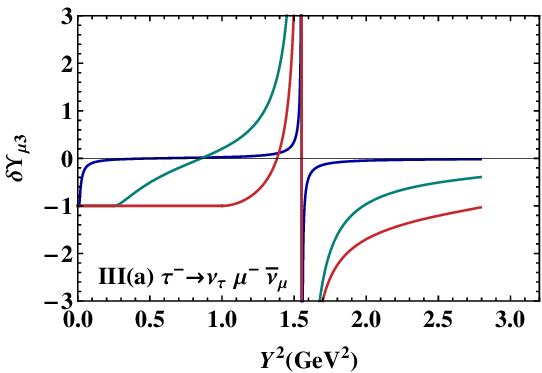}
\includegraphics[width=2.8in]{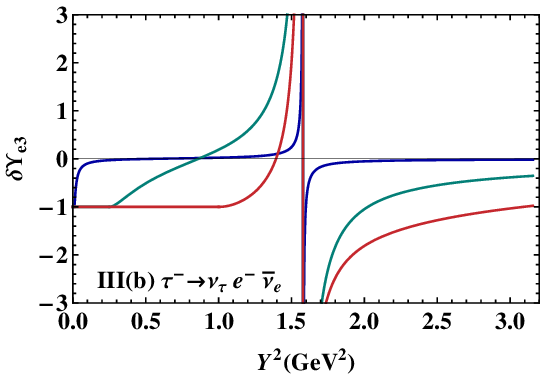}
\caption{Distributions of $\delta\Upsilon_{\ell1}$, $\delta\Upsilon_{\ell2}$,
  and $\delta\Upsilon_{\ell3}$ as functions of $Y^2$ in panels (I), (II), and
  (III), respectively. (a): $\tau^- \to \nu_\tau \mu \bar{\nu}_\mu$, (b):
  $\tau^- \to \nu_\tau e \bar{\nu}_e$. The Dark Blue, Teal, and Dark Red lines
  represent $m_{4\nu}=(0.1 ,0.5, 1.0) \text{GeV}$, respectively.
}
\label{triangle2}
\end{figure}

We consider three sterile neutrino masses: 0.1, 0.5, and 1.0~GeV. The
corresponding distributions of the parameters $\delta\Upsilon_{\ell1}$,
$\delta\Upsilon_{\ell2}$, and $\delta\Upsilon_{\ell3}$ as functions of $Y^2$
are shown in Fig.~\ref{triangle2}. These parameters include a common factor
$\frac{1}{U^2_{\ell^{\prime}4}+U^2_{\ell4}}$, whose magnitude can be
constrained by experimental observations. This, in turn, restricts the
potential influence of sterile neutrinos on the extended
Pontecorvo-Maki-Nakagawa-Sakata matrix~\cite{Maki:1962mu,Pontecorvo:1967fh}.
We find that the distributions for $\delta\Upsilon_{\ell2}$ and
$\delta\Upsilon_{\ell3}$ exhibit singularities. These arise because the
corresponding SM values for $\Upsilon^S_{\ell2}$ and $\Upsilon^S_{\ell3}$
vanish, whereas their new physics counterparts do not. The singularity occurs
near $Y^2 \simeq m^2_{\ell^{\prime}}/2$. This indicates that this behavior
manifests only for $m_{4\nu}^2 < m^2_{\ell^{\prime}}/2$, a region deserving
particular attention in experimental measurements. In other words, only sterile
neutrinos lighter than the parent charged lepton could cause observable
effects. In particular, due to the condition
$m_{4\nu}^2 < m^2_{\ell^{\prime}}/2$, the restriction from phase space makes
leptonic decays of the $\tau$ a better candidate than those of the $\mu$ for
probing sterile neutrinos.

Existing experiments have already collected an impressive number of $\tau$
lepton samples, for instance, Belle~II has accumulated approximately
$45 \times 10^9$ $\tau$-pair events with only 8\% non-$\tau$ background,
while future colliders such as CEPC and FCC are projected to deliver
$70 \times 10^9$ and $170 \times 10^9$ $\tau$ pairs, respectively, with
even cleaner conditions~\cite{yuan2021tau}. These datasets present outstanding
opportunities for precision studies of $\tau$ decays.

However, these samples are currently unpolarized. If future colliders were to
incorporate polarized $e^+e^-$ beams, they could produce polarized $\tau$
leptons, whose polarization could be reconstructed from angular distributions
relative to the beam axis. This would allow significantly more sensitive tests
of the SM and deeper probes of new physics, including the sterile neutrino
signatures proposed in this work. Our analysis therefore identifies a promising
phenomenological target and motivates the development of polarized-beam
capabilities at next-generation colliders.


\section{Summary}
We have investigated the decays
$\ell^{\prime-} \to \ell^- \bar{\nu_\ell} \nu_{\ell^{\prime}}$ and discussed
their polarization-dependent asymmetries in order to probe the possible
existence of sterile neutrinos. Our results show that sterile neutrinos can
cause singularities in the distribution of the proposed asymmetries parameters
as function of $Y^2$.

These findings motivate the inclusion of polarized beam sources at future
colliders, which would enhance sensitivity to neutrino-related new physics
via polarization observables and could potentially uncover new physics beyond
the SM.

\section*{ACKNOWLEDGMENTS}
HZ and BCK were supported in part by the Excellent Youth Foundation of Henan Scientific Committee under Contract No.~242300421044, National Natural Science Foundation of China~(NSFC) under Contract No.~12192263 and Joint Large-Scale Scientific Facility Fund of the NSFC and the Chinese Academy of Sciences under Contract No.~U2032104; YY was supported in part by National Natural Science Foundation of China~(NSFC) under Contracts No.~11905023, No.~12047564 and No.~12147102, the Natural Science Foundation of Chongqing (CQCSTC) under Contract No.~cstc2020jcyj-msxmX0555 and the Science and Technology Research Program of Chongqing Municipal Education Commission (STRPCMEC) under Contracts No.~KJQN202200605 and No.~KJQN202200621.


\begin{thebibliography}{99}

\bibitem{Davis:1968cp}
R.~Davis, Jr., D.~S.~Harmer and K.~C.~Hoffman,
Phys. Rev. Lett. \textbf{20}, 1205-1209 (1968)
doi:10.1103/PhysRevLett.20.1205

\bibitem{Super-Kamiokande:1998kpq}
Y.~Fukuda \textit{et al.} [Super-Kamiokande],
Phys. Rev. Lett. \textbf{81}, 1562-1567 (1998)
doi:10.1103/PhysRevLett.81.1562
[arXiv:hep-ex/9807003 [hep-ex]].
\bibitem{SNO:2002tuh}
Q.~R.~Ahmad \textit{et al.} [SNO],
Phys. Rev. Lett. \textbf{89}, 011301 (2002)
doi:10.1103/PhysRevLett.89.011301
[arXiv:nucl-ex/0204008 [nucl-ex]].
\bibitem{KamLAND:2002uet}
K.~Eguchi \textit{et al.} [KamLAND],
Phys. Rev. Lett. \textbf{90}, 021802 (2003)
doi:10.1103/PhysRevLett.90.021802
[arXiv:hep-ex/0212021 [hep-ex]].
\bibitem{K2K:2002icj}
M.~H.~Ahn \textit{et al.} [K2K],
Phys. Rev. Lett. \textbf{90}, 041801 (2003)
doi:10.1103/PhysRevLett.90.041801
[arXiv:hep-ex/0212007 [hep-ex]].
\bibitem{Marquez:2022bpg}
J.~M.~M{\'a}rquez, G.~L.~Castro and P.~Roig,
JHEP \textbf{11}, 117 (2022)
doi:10.1007/JHEP11(2022)117
[arXiv:2208.01715 [hep-ph]].

\bibitem{Beuthe:2001rc}
M.~Beuthe,
Phys. Rept. \textbf{375}, 105-218 (2003)
doi:10.1016/S0370-1573(02)00538-0
[arXiv:hep-ph/0109119 [hep-ph]].
\bibitem{Dasgupta:2021ies}
B.~Dasgupta and J.~Kopp,
Phys. Rept. \textbf{928}, 1-63 (2021)
doi:10.1016/j.physrep.2021.06.002
[arXiv:2106.05913 [hep-ph]].
\bibitem{Bryman:2019ssi}
D.~A.~Bryman and R.~Shrock,
Phys. Rev. D \textbf{100}, no.5, 053006 (2019)
doi:10.1103/PhysRevD.100.053006
[arXiv:1904.06787 [hep-ph]].
\bibitem{Bryman:2019bjg}
D.~A.~Bryman and R.~Shrock,
Phys. Rev. D \textbf{100}, 073011 (2019)
doi:10.1103/PhysRevD.100.073011
[arXiv:1909.11198 [hep-ph]].
\bibitem{Planck:2018vyg}
N.~Aghanim \textit{et al.} [Planck],
Astron. Astrophys. \textbf{641}, A6 (2020)
[erratum: Astron. Astrophys. \textbf{652}, C4 (2021)]
doi:10.1051/0004-6361/201833910
[arXiv:1807.06209 [astro-ph.CO]].
\bibitem{CUORE:2017tlq}
C.~Alduino \textit{et al.} [CUORE],
Phys. Rev. Lett. \textbf{120}, no.13, 132501 (2018)
doi:10.1103/PhysRevLett.120.132501
[arXiv:1710.07988 [nucl-ex]].
\bibitem{GERDA:2018pmc}
M.~Agostini \textit{et al.} [GERDA],
Phys. Rev. Lett. \textbf{120}, no.13, 132503 (2018)
doi:10.1103/PhysRevLett.120.132503
[arXiv:1803.11100 [nucl-ex]].
\bibitem{Ke:2023qzc}
B.~C.~Ke, J.~Koponen, H.~B.~Li and Y.~Zheng,
Ann. Rev. Nucl. Part. Sci. \textbf{73}, 285-314 (2023)
doi:10.1146/annurev-nucl-110222-044046
[arXiv:2310.05228 [hep-ex]].
\bibitem{Cabibbo:1965zzb} N.~Cabibbo and A.~Maksymowicz, \href{https://journals.aps.org/pr/abstract/10.1103/PhysRev.137.B438}{Phys. Rev. \textbf{137}, B438 (1965)} [erratum: \href{https://journals.aps.org/pr/abstract/10.1103/PhysRev.168.1926}{Phys. Rev. \textbf{168}, 1926 (1968)}].
\bibitem{Shrock:1981cq}
R.~E.~Shrock,
Phys. Lett. B \textbf{112}, 382-386 (1982)
doi:10.1016/0370-2693(82)91074-7
\bibitem{Shrock:1980ct}
R.~E.~Shrock,
Phys. Rev. D \textbf{24}, 1232 (1981)
doi:10.1103/PhysRevD.24.1232
\bibitem{Shrock:1981wq}
R.~E.~Shrock,
Phys. Rev. D \textbf{24}, 1275 (1981)
doi:10.1103/PhysRevD.24.1275

\bibitem{Maki:1962mu}
Z.~Maki, M.~Nakagawa and S.~Sakata,
Prog. Theor. Phys. \textbf{28}, 870-880 (1962)
\bibitem{Pontecorvo:1967fh}
B.~Pontecorvo,
Zh. Eksp. Teor. Fiz. \textbf{53}, 1717-1725 (1967)
\bibitem{yuan2021tau}
C.-Z.~Yuan, ``Particle identification for $\tau$ physics'' (slide presentation), 
IAS Program on High Energy Physics 2021, Jan.~14--21, 2021. 
URL: \url{https://indico.global/event/12243/contributions/108663/attachments/49933/95931/tau_physics_Yuan.pdf}

\end{thebibliography}
\end{document}